\documentclass{article}
\usepackage{spconf,amsmath}
\usepackage{amssymb,mathrsfs}
\usepackage{color,array,times}
\usepackage{epsfig,epstopdf}
\title{Study of Joint MSINR and Relay Selection Algorithms for Distributed
Beamforming \vspace{-0.5em}}
\name{Hang Ruan and Rodrigo C. de
Lamare \vspace{-0.5em} }
\address{Department of Electronics, University of York, UK\\
CETUC, PUC-Rio, Brazil\\
Emails: hr648@york.ac.uk and rodrigo.delamare@york.ac.uk
\vspace{-0.5em}}

\begin{document}
\ninept
\maketitle

\begin{abstract}
This paper presents joint maximum signal-to-interference-plus-noise
ratio (MSINR) and relay selection algorithms for distributed
beamforming. We propose a joint MSINR and restricted greedy search
relay selection (RGSRS) algorithm with a total relay transmit power
constraint that iteratively optimizes both the beamforming weights
at the relays nodes, maximizing the SINR at the destination.
Specifically, we devise a relay selection scheme that based on
greedy search and compare it to other schemes like restricted random
relay selection (RRRS) and restricted exhaustive search relay
selection (RESRS). A complexity analysis is provided and simulation
results show that the proposed joint MSINR and RGSRS algorithm
achieves excellent bit error rate (BER) and SINR performances.
\end{abstract}


\section{Introduction}

Distributed beamforming has been widely investigated in wireless
communications \cite{r1,r2,r3} and cooperative diversity approaches
\cite{TDS_clarke,TDS_2,switch_int,switch_mc,smce,TongW,jpais_iet,TARMO,badstc,baplnc}
in recent years. It is key for situations in which the channels
between the sources and the destination have poor quality so that
devices cannot communicate directly and the destination relies on
relays that receive and forward the signals \cite{r2}. The work in
\cite{r3} formulates an optimization problem that maximizes the
output signal-to-interference-plus-noise ratio (SINR) under the
individual relay power constraints. The work in \cite{r4,r7} focuses
on the optimization of weights using all relays to increase the SINR
in relay networks. Another related work \cite{r13} derives a
reference signal based scheme that only uses the local channel state
information (CSI).

However, in most scenarios relays are either not ideally distributed in terms
of locations or the channels involved with some of the relays have poor
quality. Possible solutions can be categorized in two approaches. One is to
adaptively adjust the power of each relay according to the qualities of its
associated channels, known as adaptive power control or power allocation. Some
power control methods based on channel magnitude and relative analysis have been
studied in \cite{r5,r6}. An alternative solution is to use relay selection, which
selects a number of relays according to a criterion of interest while
discarding the remaining relays. In \cite{r8,r14}, several optimum single-relay selection schemes and
a multi-relay selection scheme using relay ordering based on maximizing the
output SNR under individual relay power constraints are developed and
discussed, but the beamforming weights are not optimized iteratively and synchronously to enhance the SINR maximization.
The work in \cite{r9} proposed a low-cost greedy search method for
the uplink of cooperative direct sequence code-division multiple access
systems, which approaches the performance of an exhaustive search.
In \cite{r12}, multi-relay selection algorithms have been developed to maximize
the secondary receiver in a two-hop cognitive relay network. In \cite{r15},
a combined cooperative beamforming and relay selection scheme that only selects two relays is proposed for physical layer security.

In this work, we propose a joint MSINR distributed beamforming and
restricted greedy search relay selection (RGSRS) algorithm with a
total relay transmit power constraint which iteratively optimizes
both the beamforming weights
\cite{xutsa,delamaretsp,kwak,xu&liu,delamareccm,wcccm,delamareelb,jidf,delamarecl,delamaresp,delamaretvt,jioel,rrdoa,delamarespl07,delamare_ccmmswf,jidf_echo,delamaretvt10,delamaretvt2011ST,delamare10,fa10,lei09,ccmavf,lei10,jio_ccm,ccmavf,stap_jio,zhaocheng,zhaocheng2,arh_eusipco,arh_taes,dfjio,rdrab,locsme,okspme,dcg_conf,dcg,dce,drr_conf,dta_conf1,dta_conf2,dta_ls,damdc,song,wljio,barc,jiomber,saalt}
at the relay nodes, maximizing the ouput SINR at the destination,
provided that the second-order statistics of the CSI is perfectly
known. Specifically, we devise a relay selection scheme based on a
greedy search and compare it to other schemes like restricted random
relay selection (RRRS) and restricted exhaustive search relay
selection (RESRS). The RRRS scheme selects a fixed number of relays
randomly from all relays. The RESRS scheme employs the exhaustive
search method that runs every single possible combination among all
relays aiming to obtain the set with the best SINR performance. The
proposed RGSRS scheme is developed from a greedy search method with
a specific optimization problem that works in iterations and
requires SINR feedback from the destination. These joint MSINR and
restricted relay selection methods are compared with the scenario
without relay selection and the results show significant
improvements in terms of SINR and BER performances of the proposed
algorithm. The computational cost of all algorithms are analyzed.


This paper is organized as follows. Section 2 presents the system model. In Section 3, the joint MSINR and relay selection method is introduced. Section 4 derives the joint MSINR and RGSRS algorithm and provides its computational complexity analysis. Section 5 presents the simulation results. Section 6 gives the conclusion.

\section{System Model}

We consider a wireless communication network consisting of $K$
signal sources (one desired signal with the others as interferers),
$M$ distributed single-antenna relays and a destination. It is
assumed that the quality of the channels between the signal sources
and the destination is poor so that direct communications are
impossible and their links are negligible. The $M$ relays receive
information transmitted by the signal sources and then retransmit to
the destination as a beamforming procedure, in which a two-step
amplify-and-forward (AF) protocol is considered for cooperative
communications.


In the first step, the sources transmit the signals to the relays as
\begin{equation}
{\bf x}={\bf F}{\bf s}+{\boldsymbol \nu}, \label{eq1}
\end{equation}
where ${\bf s}=[s_1, s_2, \dotsb, s_K]^T \in {\mathbb C}^{K \times 1}$ are signal sources with zero mean, $[.]^T$ denotes the transpose, $s_k=\sqrt{P_{s,k}}s$, $E[|s|^2]=1$, $P_{s,k}$ is the transmit power of the $k$th signal source, $k=1,2,\dotsb,K$, $s$ is the information symbol. Without loss of generality we can assume $s_1$ as the desired signal while the others are treated as interferers. ${\bf F}=[{\bf f}_1, {\bf f}_2, \dotsb, {\bf f}_K] \in {\mathbb C}^{M \times K}$ is the channel matrix between the signal sources and the relays, ${\bf f}_k=[f_{1,k}, f_{2,k}, \dotsb, f_{M,k}]^T \in {\mathbb C}^{M \times 1}$, $f_{m,k}$ denotes the channel between the $m$th relay and the $k$th source ($m=1,2, \dotsb, M$, $k=1,2,\dotsb, K$). ${\boldsymbol \nu}=[\nu_1, \nu_2, \dotsb, \nu_M]^T \in {\mathbb C}^{M \times 1}$ is the complex Gaussian noise vector at the relays and $\sigma_{\nu}^2$ is the noise variance at each relay ($\nu_m$ \~{} ${\mathcal CN}(0,\sigma_{\nu}^2)$). The vector ${\bf x} \in {\mathbb C}^{M \times 1}$ represents the received data at the relays. In the second step, the relays transmit ${\bf y} \in {\mathbb C}^{M \times 1}$ which is an amplified and phase-steered version of ${\bf x}$ that can be written as
\begin{equation}
{\bf y}={\bf W}{\bf x}, \label{eq2}
\end{equation}
where ${\bf W}={\rm diag}([w_1, w_2, \dotsb, w_M]) \in {\mathbb
C}^{M \times M}$ is a diagonal matrix whose entries denote the
beamforming weights. Then the signal received at the destination is
given by
\begin{equation}
z={\bf g}^T{\bf y}+n, \label{eq3}
\end{equation}
where $z$ is a scalar, ${\bf g}=[g_1, g_2, \dotsb, g_M]^T \in
{\mathbb C}^{M \times 1}$ is the complex Gaussian channel vector
between the relays and the destination, $n$ ($n$ \~{} ${\mathcal
CN}(0,\sigma_n^2)$, $\sigma_n^2=\sigma_{\nu}^2$) is the noise at the
destination and $z$ is the received signal at the destination.

It should be noted that both ${\bf F}$ and ${\bf g}$ are modeled as
Rayleigh distributed. Using the Rayleigh distribution for the
channels, we also consider distance based large-scale channel
propagation effects that include fading (or path loss) and
shadowing. Distance based fading (or path loss) is a representation
of how a signal is attenuated the further it travels in the medium
the system operates within, and can be heavily affected by the
environment. \cite{r10,r11} An exponential based path loss model can
be described by
\begin{equation}
\gamma=\frac{\sqrt{L}}{\sqrt{d^{\rho}}}, \label{eq4}
\end{equation}
where $\gamma$ is the distance based path loss, $L$ is the known path loss at the destination, $d$ is the distance of interest relative to the destination and $\rho$ is the path loss exponent, which can vary due to different environments and is typically set within $2$ to $5$, with a lower value representing a clear and uncluttered environment which has a slow attenuation and a higher value describing a cluttered and highly attenuating environment. Shadow fading can be described as a random variable with a probability distribution for the case of large scale fading as
\begin{equation}
\beta=10^{(\frac{\sigma_s{\mathcal N}(0,1)}{10})}, \label{eq5}
\end{equation}
where $\beta$ is the shadowing parameter, ${\mathcal N}(0,1)$ means the Gaussian distribution with zero mean and unit variance, $\sigma_s$ is the shadowing spread in dB. The shadowing spread reflects the severity of the attenuation caused by shadowing, and is typically given between $0$dB to $9$dB.
The channels modeled with both path-loss and shadowing can therefore be represented as:
\begin{equation}
{\bf F}=\gamma\beta{\bf F}_0,  \label{eq6}
\end{equation}
\begin{equation}
{\bf g}=\gamma\beta{\bf g}_0,  \label{eq7}
\end{equation}
where ${\bf F}_0$ and ${\bf g}_0$ denote the Rayleigh distributed channels without large-scale propagation effects \cite{r10,r11}.

\section{Proposed Joint MSINR Beamforming and Relay Selection}

In many cases of relay networking, some of the relays are quite far
away from either the signal sources or the destinations, which means
they may contribute to poor network performance due to their poor
performance for receiving and transmitting signals. The aim of joint
maximum SINR beamforming and relay selection is to compute the
beamforming weights according to the maximum SINR criterion and
optimize the relay system by discarding the relays with poor
performance and making the best use of the relays with good channels
in order to improve the overall system performance.

A joint SINR maximization problem with relay selection using a total
relay transmit power constraint encountering interferers can be
generally described as
\begin{equation}
\begin{aligned}
{\mathcal S}_{\rm opt}=\underset{\boldsymbol\alpha, {\bf w}}{\rm arg~max}~~
SINR({\mathcal S},{\mathcal H},{\bf P}_s,{\bf P}_r,P_T,\boldsymbol\alpha, {\bf w}) \\
{\rm subject ~to} ~ \sum_{m=1}^M{\alpha}_m^2P_{r,m} \leq P_T, \\
\alpha_m \in \{0,1\},m=1,2,\dotsb,M  \label{eq8}
\end{aligned}
\end{equation}
where ${\mathcal S}_{\rm opt}$ is the optimum relay set of size
$M_{\rm opt}$ ($1\leq{M_{\rm opt}}\leq{M}$) and $SINR$ is a function
of ${\mathcal S}$,${\mathcal H}$,${\bf P}_s$,${\bf P}_r$ and $P_T$,
where ${\mathcal S}$ is the original relay set of size $M$,
${\mathcal H}$ is the set containing parameters of the CSI (i.e.,
${\mathcal H}=\{{\bf F},{\bf g},\sigma_{\nu}^2\}$), ${\bf
P}_s=[P_{s,1}, P_{s,2}, \dotsb, P_{s,K}] \in {\mathbb R}^{1 \times
K}$, $k=1,2,\dotsb,K$, ${\bf P}_r=[P_{r,1}, P_{r,2}, \dotsb,
P_{r,M}]^T \in {\mathbb R}^{M \times 1}$, $m=1,2,\dotsb,M$,
$P_{r,m}$ refers to the transmit power of the $m$th relay (Note that
before selection we have $\sum_{m=1}^MP_{r,m}\leq{P_T}$ and we
consider that each relay cooperates with its full power as long as
it is selected), $P_T$ is the maximum allowable total transmit power
of all relays, ${\boldsymbol\alpha}=[\alpha_1, \alpha_2, \dotsb,
\alpha_M]^T$, $\alpha_m$ ($m=1,\dotsb,M$) is the relay cooperation
parameter which determines wether the $m$th relay cooperates or not,
${\bf w}=[w_1, w_2, \dotsb, w_M]^T \in {\mathbb C}^{M \times 1}$ is
the beamforming weight vector. The received signal at the $m$th
relay as:
\begin{equation}
x_m=\sum_{k=1}^K\sqrt{P_{s,k}}{s}f_{m,k}+{\nu}_m, \label{eq9}
\end{equation}
then the transmitted signal at the $m$th relay can be written as:
\begin{equation}
y_m=\alpha_m{w_m}x_m. \label{eq10}
\end{equation}
Note that we can express the transmit power at the $m$th relay
$P_{r,m}$ as $E[|y_m|^2]$ so that the total relay transmit power can
be written as
$\sum_{m=1}^ME[|y_m|^2]=\sum_{m=1}^ME[|\alpha_m{w_m}x_m|^2]$ or in
matrix form as $(\boldsymbol\alpha^H\odot{\bf w}^H){\bf
D}(\boldsymbol\alpha\odot{\bf w})$ where ${\bf D}={\rm
diag}(\boldsymbol\alpha\odot(\sum_{k=1}^KP_{s,k}[E[|f_{1,k}|^2],
E[f_{2,k}|^2], \dotsb, E[f_{M,k}|^2]])+\sigma_n^2)$ is a full-rank
matrix. The signal received at the destination can be expanded by
substituting \eqref{eq9} and \eqref{eq10} in \eqref{eq7}, which
gives
\begin{multline}
z=\underbrace{\sum_{m=1}^M\alpha_m{w_m}g_m\sqrt{P_{s,1}}f_{m,1}s}_{\text{desired
signal}}
+\underbrace{\sum_{m=1}^M\alpha_m{w_m}g_m\sum_{k=2}^K\sqrt{P_{s,k}}f_{m,k}s}_{\text{interferers}}\\
+\underbrace{\sum_{m=1}^M\alpha_m{w_m}g_m\nu_m+n}_{\text{noise}}.
\label{eq11}
\end{multline}
By taking expectations of the components of \eqref{eq11}, we can
compute the desired signal power $P_{z,1}$, the interference power
$P_{z,i}$ and the noise power $P_{z,n}$ at the destination as
follows:
\begin{multline}
P_{z,1}=E[\sum_{m=1}^M(\alpha_m{w_m}g_m\sqrt{P_{s,1}}f_{m,1}s)^2] \\
=P_{s,1}\sum_{m=1}^M\alpha_m^2E[w_m^*(f_{m,1}g_m)(f_{m,1}g_m)^*w_m], \label{eq12}
\end{multline}
\vspace{-1.5em}
\begin{multline}
P_{z,i}=E[(\sum_{m=1}^M(\alpha_m{w_m}g_m\sum_{k=2}^K\sqrt{P_{s,k}}f_{m,k}s)^2] \\
=\sigma_n^2(1+\alpha_m^2\sum_{m=1}^ME[w_m^*g_mg_m^*w_m]), \label{eq13}
\end{multline}
\vspace{-1.5em}
\begin{multline}
P_{z,n}=E[\sum_{m=1}^M(\alpha_m{w_m}g_m\nu_m+n)^2] \\
=\sum_{k=2}^KP_{s,k}\sum_{m=1}^M\alpha_m^2E[w_m^*(f_{m,k}g_m)(f_{m,k}g_m)^*w_m],
\label{eq14}
\end{multline}
where $*$ denotes complex conjugation. The SINR is computed as:
\begin{multline}
SINR=\frac{P_{z,1}}{P_{z,i}+P_{z,n}}. \\
=\frac{(\boldsymbol\alpha^H\odot{\bf w}^H){\bf
R}_1(\boldsymbol\alpha\odot{\bf w})}{\sigma_n^2
+(\boldsymbol\alpha^H\odot{\bf w}^H)({\bf Q}+\sum_{k=2}^K{\bf
R}_k)(\boldsymbol\alpha\odot{\bf w})}. \label{eq16}
\end{multline}

\subsection{{Computation of Weights}}

By defining $\boldsymbol\alpha\odot{\bf w}=\tilde{\bf w}$ and therefore the original problem \eqref{eq8} can be cast in terms of solving for $\tilde{\bf w}$ as:
\begin{equation}
\begin{aligned}
\underset{\tilde{\bf w}}{\rm max}~~ \frac{\tilde{\bf w}^H{\bf R}_1\tilde{\bf w}}{\sigma_n^2+\tilde{\bf w}^H({\bf Q}+\sum_{k=2}^K{\bf R}_k)\tilde{\bf w}} \\
s.t. ~~~~ \tilde{\bf w}^H{\bf D}\tilde{\bf w} \leq P_T, \\
{\rm Rank}(\tilde{\bf w}\tilde{\bf w}^H)={\rm Rank}(\boldsymbol\alpha\boldsymbol\alpha^H),  \\
\alpha_m \in \{0,1\},m=1,2,\dotsb,M,  \label{eq17} \\
\end{aligned}
\end{equation}
where ${\bf R}_1$, ${\bf Q}$ and ${\bf R}_k$ are covariance matrices
that are associated with the desired signal, the noise at the relays
and the $k$th interferer and defined by
$P_{s,1}E[(\boldsymbol\alpha\odot{\bf f}_1\odot{\bf
g})(\boldsymbol\alpha\odot{\bf f}_1\odot{\bf g})^H] \in {\mathbb
C}^{M \times M}$, $\sigma_n^2E[(\boldsymbol\alpha\odot{\bf
g})(\boldsymbol\alpha\odot{\bf g})^H] \in {\mathbb C}^{M \times M}$,
$P_{s,k}E[(\boldsymbol\alpha\odot{\bf f}_k\odot{\bf
g})(\boldsymbol\alpha\odot{\bf f}_k\odot{\bf g})^H] \in {\mathbb
C}^{M \times M}$, respectively, where $\odot$ denotes the
Schur-Hadamard product which computes element-wise multiplications,
they are such defined so that their ranks are equal to the number of
non-zero elements of $\boldsymbol\alpha$. The second constraint
indicates and ensures $\tilde{\bf w}$ has the same number of zero
elements as $\boldsymbol\alpha$ and ${\rm Rank}$ denotes the rank
operator. At this point, we use an alternating optimization strategy
to obtain the solutions for both ${\bf w}$ and $\boldsymbol\alpha$,
i.e., we fix the vector parameter ${\bf w}$ and optimize
$\boldsymbol\alpha$ and vice-versa in an alternating fashion. The
number of iterations of this alternating optimization depends on
both the minimum required number of relays, which is a user defined
parameter, and if the maximum SINR is achieved, which is determined
by the system feedback. The problem in \eqref{eq17} can be solved
with respect to ${\bf w}$ in a closed-form solution as in the total
power constraint SNR maximization problem similarly to \cite{r4},
with the assumption that the second-order statistics of the CSI
(i.e., ${\mathcal H}$) is perfectly known. Then, a closed-form
solution for $\tilde{\bf w}$ is obtained by
\begin{equation}
\tilde{\bf w}=\sqrt{P_T}{\bf D}^{-\frac{1}{2}}{\mathcal P}\{{\bf E}\}, \label{eq18}
\end{equation}
and the corresponding SINR is
\begin{equation}
SINR=P_T\lambda_{max}\{{\bf E}\}, \label{eq19}
\end{equation}
where ${\mathcal P}\{.\}$ denotes the principal eigenvector operator, $\lambda_{max}\{.\}$ denotes the largest eigenvalue of the argument, ${\bf E}=(\sigma_n^2{\bf I}+P_T{\bf D}^{-\frac{1}{2}}({\bf Q}+\sum_{k=2}^K{\bf R}_k){\bf D}^{-\frac{1}{2}})^{-1}{\bf D}^{-\frac{1}{2}}{\bf R}_1{\bf D}^{-\frac{1}{2}}$ has the same rank as ${\bf R}_1$. It is easy to observe that once we know $\boldsymbol\alpha$, we can compute the optimum weights and SINR from \eqref{eq18} and \eqref{eq19}, respectively, by using only the currently selected relay nodes and their weights. The weight optimization steps are detailed in Table. \ref{table1}.

\subsection{{Relay Selection}}

In order to solve the problem in \eqref{eq17} with respect to
$\boldsymbol\alpha$, we consider
\begin{equation}
\begin{aligned}
\underset{\boldsymbol\alpha}{\rm max} ~~ {SINR} \\
{\rm subject ~to} ~~~~ \sum_{m=1}^M{\alpha}_m^2P_{r,m} \leq P_T, \\
\alpha_m \in \{0,1\},m=1,2,\dotsb,M  \label{eq20} \\
\end{aligned}
\end{equation}
that can be solved with algorithms like greedy search and exhaustive search, which can be determined by the designer. To emphasize, $\boldsymbol\alpha$ is obtained before ${\bf w}$ is computed in each recursion. An alternative way that computes ${\bf w}$ before obtaining $\boldsymbol\alpha$ also works but the above equations will be different. This joint MSINR beamforming and relay selection method requires output SINR comparisons and feedback from the destination to the relay nodes as a form of information exchange, which is similar to \cite{r8}, but weight optimization is neglected in their work.

\begin{table}
\centering
\caption{Beamforming weight vector optimization} 
\begin{tabular}{l}
\hline
1)~~Using a search algorithm to optimize and obtain $\boldsymbol\alpha$. \\
With $\boldsymbol\alpha$ and the CSI, compute the following quantities \\ using the selected relay nodes in each iteration: \\
The desired signal related covariance matrix : \\
2)~~${\bf R}_1=P_{s,1}E[(\boldsymbol\alpha\odot{\bf f}_1\odot{\bf g})(\boldsymbol\alpha\odot{\bf f}_1\odot{\bf g})^H]$ \\
The interferers related covariance matrices for $k=2,\dotsb,K$: \\
3)~~${\bf R}_k=P_{s,k}E[(\boldsymbol\alpha\odot{\bf f}_k\odot{\bf g})(\boldsymbol\alpha\odot{\bf f}_k\odot{\bf g})^H] \in {\mathbb C}^{M \times M}$ \\
The noise related covariance matrix: \\
4)~~${\bf Q}=\sigma_n^2E[(\boldsymbol\alpha\odot{\bf g})(\boldsymbol\alpha\odot{\bf g})^H]$ \\
The transmit power related full-rank matrix ${\bf D}$: \\
5)~~${\bf D}={\rm diag}(\boldsymbol\alpha\odot(\sum_{k=1}^KP_{s,k}$ \\ $[E[|f_{1,k}|^2], E[f_{2,k}|^2], \dotsb, E[f_{M,k}|^2]])+\sigma_n^2)$ \\
The defined matrix ${\bf E}$: \\
6)~~${\bf E}=(\sigma_n^2{\bf I}+P_T{\bf D}^{-\frac{1}{2}}({\bf Q}+\sum_{k=2}^K{\bf R}_k){\bf D}^{-\frac{1}{2}})^{-1}{\bf D}^{-\frac{1}{2}}{\bf R}_1{\bf D}^{-\frac{1}{2}}$ \\
Optimize and obtain the beamforming weight vector $\tilde{\bf w}$: \\
7)~~$\tilde{\bf w}=\sqrt{P_T}{\bf D}^{-\frac{1}{2}}{\mathcal P}\{{\bf E}\}$ \\
Compute the output SINR at the destination: \\
8)~~$SINR=P_T\lambda_{max}\{{\bf E}\}$ \\
\hline
\end{tabular} \label{table1}
\end{table}

\section{Proposed Joint MSINR and RGSRS Algorithm}

The joint MSINR and RGSRS algorithm works in iterations with excellent performance. We consider a user-defined parameter $M_{min}$ as a restriction to the minimum number of relays that must be used to allow a higher flexibility for the users to control the number of relays. Before the first iteration all relays are considered (i.e., ${\mathcal S}(0)={\mathcal S}$). Consequently, we solve the following problem once for each iteration in order to cancel the relay with worst performance from set ${\mathcal S}(i-1)$ and evaluate $SINR(i)$:
\begin{equation}
\begin{aligned}
{\mathcal S}(i)=\underset{\boldsymbol\alpha(i)}{\rm arg~max} ~~ SINR(i) \\
{\rm subject ~to} \sum_{m=1}^{M}{\alpha}_m^2(i)P_{r,m}(i) \leq P_T, \\
\alpha_m(i) \in \{0,1\}, \\
||\boldsymbol\alpha(i)||_1=M-i, \\
||\boldsymbol\alpha(i)-\boldsymbol\alpha(i-1)||_1=1, \\
M-i\geq{M_{min}}, \lrcorner \label{eq22}
\end{aligned}
\end{equation}
where $SINR(i)=SINR({\mathcal S}(i-1),{\mathcal H},{\bf P}_s,{\bf P}_r(i-1),P_T)$ and can be computed by \eqref{eq19}. If the SINR in the current iteration is higher than that in the previous iteration (i.e. $SINR(i)>SINR(i-1)$), then the selection process continues; if $SINR(i)\leq SINR(i-1)$, we cancel the selection of the current iteration and remain the relay set ${\mathcal S}(i-1)$ and $SINR(i-1)$. The joint MSINR and RGSRS algorithm can be implemented as in Table. \ref{table5}.

\begin{table}
\centering
\caption{Joint MSINR and RGSRS Algorithm}
\begin{tabular}{l}
\hline
{\bf step~1}: Initialize ${\mathcal S}_{opt}={\mathcal S}(0)$, $\boldsymbol\alpha(0)={\bf 1}$ and obtain \\ $SINR_{opt}=SINR(0)$ using Table. \ref{table1}. \\
{\bf step~2}: \\
for $i=1,\dotsb,M-M_{min}$ \\
~~~~solve the optimization problem \eqref{eq22} to obtain $\boldsymbol\alpha(i)$,  \\
~~~~${\mathcal S}(i)$ and compute $SINR(i)$ using Table. \ref{table1}. \\
~~~~compare $SINR(i)$ to $SINR(i-1)$, \\
~~~~if $SINR(i)>SINR(i-1)$ \\
~~~~~~~~update ${\mathcal S}_{opt}={\mathcal S}(i)$ and $SINR_{opt}=SINR(i)$. \\
~~~~else \\
~~~~~~~~keep ${\mathcal S}_{opt}={\mathcal S}(i-1)$ and $SINR_{opt}=SINR(i-1)$. \\
~~~~~~~~break. \\
~~~~end if. \\
end for. \\
\hline
\end{tabular} \label{table5}
\end{table}

At this point, we analyze the computational complexity required by the relay selection algorithms. The MSINR based method for SINR driven beamforming weights optimization has a cost of ${\mathcal O}(M^3)$ since matrix inversion and eigen-decomposition are required. However, $M$ is usually not large so that much same attentions should be paid to the computational cost caused by the number of iterations required in these relay selection algorithms. For the joint MSINR and RRRS algorithm, there is no weight vector or relay selection vector optimization required, which means there is only one iteration and the complexity is simply ${\mathcal O}(M^3)$. The joint MSINR and RESRS algorithm has the highest computational cost due to the fact it almost searches for all possible combinations of the relays even though an extra restriction of the minimum number of relays required is added in our case. With a restriction of that at least $M_{min}$ relays must be selected, the number of iterations is $\sum_{c=M_{min}}^{M}\frac{M!}{(M-c)!c!}$. In the joint MSINR and RGSRS algorithm, \eqref{eq22} is solved once per iteration, which can be done by disabling only one relay while enabling all the others and computing and comparing their output SINRs. The total number of iterations is no greater than $\frac{(2M-i+1)i}{2})$. The proposed joint MSINR and RGSRS algorithm has much lower complexity compared to the joint MSINR and RESRS algorithm when the value of $M$ is large.

\begin{table}
\small
\centering
\caption{Complexity Comparison}
\begin{tabular}{|c|c|}
\hline
Algorithms & Computational Cost \\
\hline
Joint MSINR and RRRS & ${\mathcal O}(M^3)$ \\
\hline
Joint MSINR and RESRS & $\sum_{c=M_{min}}^{M}\frac{M!}{(M-c)!c!}~{\mathcal O}(M^3)$ \\
\hline
Joint MSINR and RGSRS & $\leq \frac{(2M-i+1)i}{2}~{\mathcal O}(M^3)$ \\
\hline
\end{tabular} \label{table6}
\end{table}

\section{Simulation Results}

\begin{figure}[!htb]
\centering
\def\epsfsize#1#2{1\columnwidth}
\epsfbox{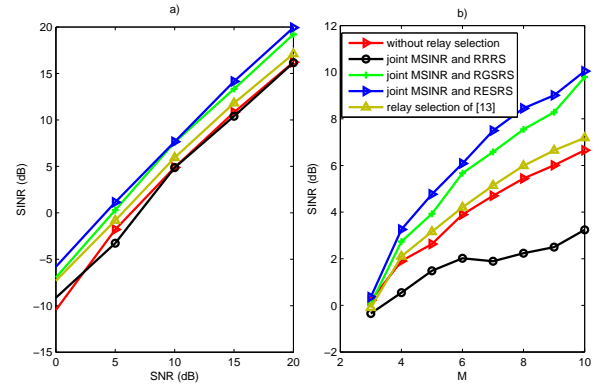}\caption{SINR versus SNR and M} \label{newfigure}
\end{figure}

\begin{figure}[!htb]
\centering
\def\epsfsize#1#2{1\columnwidth}
\epsfbox{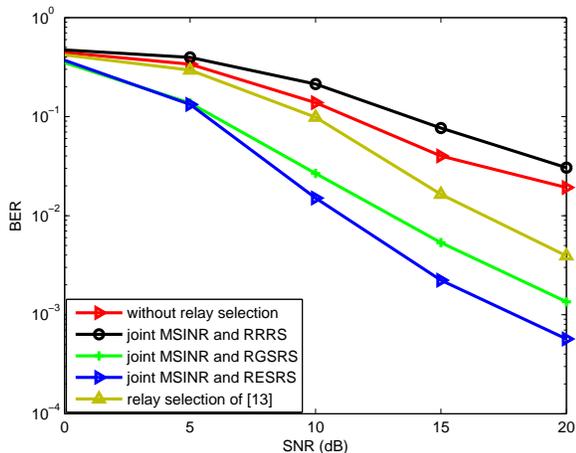}\caption{BER versus SNR}
\label{fig3}
\end{figure}

In the simulations, we compare the joint MSINR and relay selection
algorithms to the scenario without relay selection in terms of their
SINR and bit error rate (BER) performances. The generic parameters
used for all scenarios include: number of signal sources $K=3$, the
path loss exponent $\rho=2$, the power path loss from signals to the
destination $L=10$dB, shadowing spread $\sigma_s=3$dB, $P_T=1$dBW.
Fig. \ref{newfigure}-a illustrates the SINR versus SNR (from $0$dB
to $20$dB) performance of the compared algorithms, in which the
total number of relays and interference-to-noise ratio (INR) are
fixed at $M=8$ and INR=$10$dB, respectively. Fig. \ref{newfigure}-b
illustrates how the SINR varies when the total number of relays in
the network increases, in which the input SNR=$10$dB and INR=$10$dB
are fixed. In this case, a minimum total number of relays observed
is chosen as $M=3$, whereas the maximum is at $M=10$. For each of
the above two scenarios, $500$ repetitions are carried out for each
algorithm. In Fig. \ref{fig3}, we evaluate the BER versus SNR
performance of all algorithms using Binary Phase Shift Keying (BPSK)
for the system and test all algorithms with $100000$ bits, while
keeping INR=$10$dB. For all the above scenarios, we fix the number
of randomly selected relays at $3$ for the joint MSINR and random
relay selection algorithm, the minimum required selected relays also
at $3$ for the other algorithms. As observed, the joint MSINR and
RESRS and the joint MSINR and RGSRS algorithms have the best
performance.

\section{Conclusion}

We have proposed a joint MSINR and RGSRS algorithm for distributed
beamforming which is derived based on a greedy search relay
selection scheme. The computational cost of the proposed algorithm
has been analyzed and compared to prior work that employ RRRS and
RESRS schemes. The results have shown excellent SINR and BER
performances of the proposed algorithm which are very close to the
joint MSINR and RESRS algorithm.

\end{document}